\documentstyle[epsf,seceq,onecolumn]{jpsj_sissa}

\newcommand{\icom}{{\rm i}}
\newcommand{\smint}{\!\int\!}
\newcommand{\smintq}{\smint {\rm d}^2{\mib q}}

\newcommand{\Ima}{{\rm Im}}

\newcommand{\xis}{\xi_\sigma}
\newcommand{\xid}{\xi_{\rm d}}
\newcommand{\sdxi}{\xi_{\sigma,{\rm d}}}
\newcommand{\zxi}{\xi^{(0)}}
\newcommand{\zxis}{\zxi_\sigma}
\newcommand{\zxid}{\zxi_{\rm d}}
\newcommand{\zsdxi}{\zxi_{\sigma,{\rm d}}}
\newcommand{\sgamma}{\gamma_\sigma}
\newcommand{\dgamma}{\gamma_{\rm d}}
\newcommand{\sdgamma}{\gamma_{\sigma,{\rm d}}}
\newcommand{\zgamma}{\gamma^{(0)}}
\newcommand{\zsgamma}{\zgamma_\sigma}
\newcommand{\zdgamma}{\zgamma_{\rm d}}
\newcommand{\zsdgamma}{\zgamma_{\sigma,{\rm d}}}

\newcommand{\TPG}{T_{\rm PG}}

\newcommand{\Tc}{T_{\rm c}}

\newcommand{\TKT}{T_{\rm KT}}

\newcommand{\sphi}{{\mib \phi}_\sigma}
\newcommand{\dphi}{\phi_{\rm d}}
\newcommand{\bdphi}{\bar{\phi}_{\rm d}}
\newcommand{\sGamma}{{\mit \Gamma_\sigma}}
\newcommand{\dGamma}{{\mit \Gamma_{\rm d}}}
\newcommand{\schi}{\chi_\sigma}
\newcommand{\dchi}{\chi_{\rm d}}

\newcommand{\uss}{u_{\sigma\sigma}}
\newcommand{\usd}{u_{\sigma {\rm d}}}
\newcommand{\udd}{u_{\rm dd}}

\newcommand{\cs}{c_\sigma}
\newcommand{\cd}{c_{\rm d}}
\newcommand{\TtG}{T_{2{\rm G}}}

\newcommand{\As}{A_\sigma}
\newcommand{\Ad}{A_{\rm d}}

\def\runtitle{Single-Particle Pseudogap in Two-Dimensional Electron Systems}
\def\runauthor{Shigeki {\sc Onoda} and Masatoshi {\sc Imada}}

\title{Single-Particle Pseudogap in Two-Dimensional Electron Systems}

\author{Shigeki {\sc Onoda}~\footnote{E-mail: onoda@ginnan.issp.u-tokyo.ac.jp}
and Masatoshi {\sc Imada}}
\inst{ Institute for Solid State Physics, University of Tokyo, Roppongi,
Minato-ku, Tokyo 106-8666}
\recdate{}

\abst{We investigate pseudogap phenomena in the 2D electron system.
  Based on the mode-mode coupling theory of antiferromagnetic (AFM) and 
  $d_{x^2-y^2}$-wave superconducting ($d$SC) fluctuations,
  single-particle dynamics is analyzed. For the parameter values of
  underdoped cuprates, pseudogap structure grows in the
  single-particle spectral weight $A(k,\omega)$ around the wave vector 
  $(\pi,0)$ and $(0,\pi)$ below the pseudo-spin-gap temperature $\TPG$
  signaled by the reduction of dynamical spin correlations in
  qualitative agreement with the experimental data.
  The calculated results for the overdoped cuprates also reproduce the 
  absence of the pseudogap in the experiments.
  We also discuss limitations of our weak-coupling approach.
}
\kword{high-$\Tc$ superconductivity, pseudogap, pairing fluctuations,
spin fluctuations}

\begin{document}
\sloppy
\maketitle

\section{Introduction} \label{SECTION_intro}

It is still controversial how the unusual normal state,
particularly the pseudogap in the underdoped region of the high-$\Tc$
superconductors~\cite{BednorzMuller86}, is understood,
although various experimental studies suggest that
the pseudogap develops below $\TPG$ well
above the transition temperature $\Tc$~\cite{RMP,AFM-dSC}. 

ARPES (angle-resolved photoemission spectra) results in the underdoped
cuprates~\cite{LoeserShenDessau1996_Phys.Rep._Science,photoemission}
suggest the followings: 
The quasiparticle dispersion is very flat and strongly damped
around $(\pi,0)$ and $(0,\pi)$ points (``flat spots'').
The pseudogap in the single-particle excitations develops first
around the flat spots below $\TPG$,
gradually extends in the direction to the $d_{x^2-y^2}$-wave gap nodes
and seems to continuously merge into the $d$SC gap below $\Tc$.
These experimental facts suggest that fermions in the region
around the flat spots (``flat shoal region'')
are particularly important in considering the underdoped cuprates. 

NMR experiments have revealed that
the spin-lattice relaxation rate $1/{}^{63}T_1T$
has a peak around $T\sim\TPG$, while $1/\TtG$ increases down to $\Tc$, 
in many underdoped cuprates~\cite{Yasuoka,Y124NMR,Bi2212NMR}.
Neutron scattering experiments have clarified that the resonance peak
at a finite energy $\omega^*$ grows below $\TPG$, where $\omega^*$ is
smaller for smaller doping concentrations~\cite{Fong}.
These results imply that though AFM fluctuations increase down to $\TPG$,
they are suppressed around $\omega=0$,
while enhanced around $\omega^*$ below $\TPG$.

Recently, we have studied the spin pseudogap emerging
from the dominance of the $d$SC short-ranged order (SRO) over the AFM SRO
by considering the AFM and $d$SC modes on an equal footing,
their mode-mode couplings and feedback effects on the mode dampings.
It has succeeded in reproducing the magnetic properties
in the high-$\Tc$ cuprates.
In this letter using the same parameter values as Ref.~\cite{AFM-dSC},
we calculate the electronic spectra from the
previously obtained susceptibilities~\cite{AFM-dSC}.
Our results qualitatively reproduce
and the pseudogap developing around the flat
spots~\cite{LoeserShenDessau1996_Phys.Rep._Science,photoemission}.
The results also reproduce
the flat and damped dispersion around the flat spots
observed in cuprates~\cite{LoeserShenDessau1996_Phys.Rep._Science,Gofron},
and is reminiscent of numerical results for doped Mott insulators~\cite{Imada98,Assaad98}.

\section{Properties of AFM and $d$SC Fluctuations}

In the previous paper~\cite{AFM-dSC},
starting from the 2D electron system with AFM and $d$SC fluctuations,
we have considered the effective action for AFM and $d$SC collective
modes; $S=S^{(0)}+S^{(2)}+S_{\sigma\sigma}^{(4)}
+S_{\rm dd}^{(4)}+S_{\sigma {\rm d}}^{(4)}$ with
\begin{subequations}
\begin{eqnarray}
\lefteqn{\hspace{-16mm}S^{(2)}=\frac1T\sum_q\left[\schi^{-1}\!(q)|\sphi(q)|^2
+\dchi^{-1}\!(q)\bdphi(q)\dphi(q)\right]\!,}
\label{S2} \\
&&\hspace{-16mm}S_{\sigma\sigma}^{(4)}=\frac{\uss}{T}\hspace{-5pt}\sum_{q_1,q_2,q_3}
\hspace{-5pt}\sphi(q_1)\!\cdot\!\sphi(q_2)\sphi(q_3)\!\cdot\!\sphi(q_4),
\label{S_sigma4} \\
&&\hspace{-16mm}S_{\rm dd}^{(4)}=\frac{\udd}{T}\hspace{-5pt}\sum_{q_1,q_2,q_3}
\hspace{-5pt}\bdphi(q_1)\dphi(q_2)\bdphi(q_3)\dphi(q_4),
\label{S_d4} \\
&&\hspace{-16mm}S_{\sigma {\rm d}}^{(4)}=\frac{2\usd}{T}\hspace{-5pt}\sum_{q_1,q_2,q_3}
\hspace{-5pt}\sphi(q_1)\!\cdot\!\sphi(q_2)\bdphi(q_3)\dphi(q_4),
\label{S_sigmad4}
\end{eqnarray}
\end{subequations}
where $\sphi$ and $\dphi$ ($\bdphi$) are the auxilary fields for
spins and $d$-wave pairs, $q$ and $\sum_q$ represents
$(\icom\omega_n,{\mib q})$ and $\sum_n\smintq$,
$\omega_n=2\pi nT$, and $q_4=-q_1-q_2-q_3$. Here,
\begin{subequations}
\begin{eqnarray}
\lefteqn{\hspace{-9.5mm}\schi(\icom\omega_n,{\mib q})\!=\!\As\left(\zxis{}^{-2}+\tilde{\mib q}^2+\frac{\sgamma|\omega_n|}{\cs^2}+\frac{\omega_n^2}{\cs^2}\right)^{-1}\hspace*{-2mm},}
\label{chi_sigma} \\
&&\hspace{-9.5mm}\dchi(\icom\omega_n,{\mib q})\!=\!\Ad\left(\zxid{}^{-2}+{\mib q}^2+\frac{\dgamma|\omega_n|}{\cd^2}
+\frac{\omega_n^2}{\cd^2}\right)^{-1}\hspace*{-2mm},
\label{chi_d}
\end{eqnarray}
\end{subequations}
are AFM and $d$SC susceptibilities,
where $\tilde{\mib q}={\mib q}-{\mib Q}$ with ${\mib Q}=(\pi,\pi)$.
We have neglected the phase excitations
and possible long-range features of the Coulomb repulsion
which may lead to gapful $d$SC excitations.
It can be justified because they do not seem to alter the pseudogap
structure produced by the growth of pairing amplitude.
Long-range features of Coulomb repulsion
which may lead to the gapful phase modes even in the $d$SC phase
do not alter the following main argument.
$\cs$ ($\cd$) and $\dgamma$ ($\sgamma$) are the velocity and the
damping constant of spin (pairing) modes.
The spin and the $d$SC correlation length $\zxis$ and $\zxid$ are given by
$\zxis{}^{-2}\approx1-\frac{|\sGamma|}{t}\log\frac{t}{max\{\mu,t',T\}}
\log\frac{t}{max\{\mu,T\}}$ and
$\zxid{}^{-2}\approx1-\frac{|\dGamma|}{\sqrt{t^2-4t'^2}}
\log\frac{t}{T}\log\frac{t}{max\{\mu,T\}}$, respectively,
within the Gaussian approximation
for only the dominant flat-spot contributions.
Using the transfers for the nearest-neighbor $t$ and the second-neighbor $t'$
and the chemical potential $\mu$ measured from $(\pi,0)$,
the bare dispersion is given by
$\varepsilon({\bf k})=-2t(\cos k_x+\cos k_y)-4t'(\cos k_x\cos k_y+1)-\mu$. 
Coupling of fermions to spins is $\sGamma$ and to $d$-wave pairs is $\dGamma$.
We neglect the $t'$-, $\mu$- and $T$-dependences of the other
parameters in (2.1) than $\zsdxi{}^{-2}$.

To include feedback effects from the growth of $\sdxi$ on $\sdgamma$, we take
\begin{equation}
\sdgamma=2\zsdgamma/(\xis^\varphi+\xid^\varphi)
\label{damp_z=1whole}
\end{equation}
from phenomenological arguments~\cite{AFM-dSC}.
If the flat-spot fermions dominate the dampings, we take $\varphi=1$.
When other fermions dominate the dampings, $\varphi=0$.

The AFM-AFM, $d$SC-$d$SC and AFM-$d$SC couplings, $\uss$, $\udd$ and $\usd$,
respectively, are assumed to be positive for cuprates~\cite{AFM-dSC},
where AFM and $d$SC fluctuations compete at low energies.

We have calculated the spin and $d$SC correlation lengths $\xis$
and $\xid$ by the self-consistent renormalization (SCR).
Though the SCR in 2D can not describe the Kosterlitz-Thouless
transition at $T=\TKT$,
we regard the transition at $\TKT\sim T_*$ below which $\xid$ rapidly grows and $\xis$ starts decreasing.

For YBa${}_2$Cu${}_3$O${}_y$, we have estimated $t=0.25$eV from
ARPES~\cite{MassiddaYuFreeman87-1,SiZhaLevinLu93,YBCO1236.9ARPES},
$\cs=0.5t$, $\zsgamma=t$ and $\As=2t^{-1}$
from NMR and neutron scattering experiments~\cite{Y1236.69neutron,Y123NMR,Harashina94,Mook98}.
For simplicity, we take $\cd=\cs$ and $\zdgamma=\zsgamma$.
We take $\sGamma=\dGamma$,
$|t'|\sim\mu\sim\Tc$, $\uss$, $\udd$ and $\usd$ as adjustable parameters,
and choose them to give reasonable values for $T_*\sim\Tc$,
$\xis(T_*)$ and $\TPG$ compared with observed values.
We have qualitatively reproduced the overall magnetic properties observed
in YBa${}_2$Cu${}_3$O${}_y$ with $y=7$ and $6.63$, for $\varphi=0$ and $1$,
respectively~\cite{AFM-dSC,Yasuoka}.

\section{Pseudogap in Single-Particle Excitations}

We show that the previously obtained AFM and $d$SC susceptibilities
reproduce the low-energy properties of the single-particle excitations
in high-$\Tc$ cuprates as well.
Although we have reproduced the magnetic properties
in YBa${}_2$Cu${}_3$O${}_y$,
it is difficult to discuss the single-particle excitations in the same 
compounds, because the interpretation of the ARPES
data~\cite{YBCO1236.9ARPES,YBCO1236.5APRPES}
is still controversial due to the three-dimensionality,
the contributions from CuO chains, lack of evidence for the $d$SC gap,
and surface problems.~\cite{LoeserShenDessau1996_Phys.Rep._Science}
Here we compare our results with ARPES data in
Bi2212 with similar values for $\TPG(\sim 170{\rm K})$
and $\Tc(\sim 83{\rm K})$ to those in YBa${}_2$Cu${}_3$O${}_{6.63}$.

We calculate the electronic spectra $\Ima G(\omega,{\mib k})$ using the same parameter values as Ref.~\cite{AFM-dSC}.
Here, $G(\omega,{\mib k})=
1/(\omega-\varepsilon({\mib k})-\Sigma(\omega,{\mib k}))$
is the dressed Green's function.
Unfortunately, there is no available formalism to reliably calculate
the self-energy $\Sigma(\omega,{\mib k})$ or $G(\omega,{\mib k})$
in a self-consistent manner in the pseudogap region.
A similar difficulty was already pointed out by Vilk and Tremblay
in the context of the AFM fluctuations
in the 2D Hubbard model~\cite{VilkTremblay97}.
Here we calculate the self-energy within the 1-loop level using
\begin{eqnarray}
\lefteqn{\Ima\Sigma(\omega,{\mib k})=
\smint\frac{{\rm d}^2{\mib k}'}{(2\pi)^2}
\int\!\frac{{\rm d}\omega'}{2\pi}\Ima G^{(0)}(\omega',{\mib k}')}
\nonumber\\
&&\times\left[\sGamma^2\Ima\schi(\omega-\omega',{\mib k}-{\mib k}')
\left(\coth\frac{\omega-\omega'}{2T}+\tanh\frac{\omega'}{2T}\right)\right.
\nonumber\\
&&{}+\dGamma^2\frac{g({\mib k})^2+g({\mib k}')^2}{2}
\Ima\dchi(\omega+\omega',{\mib k}+{\mib k}')
\nonumber\\
&&\times\left.\left(\coth\frac{\omega+\omega'}{2T}-\tanh\frac{\omega'}{2T}\right)\right],
\label{Sigma}
\end{eqnarray}
with the bare Green's function $G^{(0)}(\omega,{\mib k})$
and $g({\mib k})=(\cos k_x-\cos k_y)/2$.
Here $\zsdxi{}^{-2}$ in (2.2) has been replaced with $\sdxi^{-2}$.
For the prefactor $\Ad$ in (\ref{chi_sigma}),
we take $\Ad=4t^{-1}$ to give a proper value for the
midpoint shift
in ARPES intensity in the pseudogap region~\cite{photoemission}.

\begin{figure}[tb]
\begin{center}
\epsfxsize=7.8cm
$$\epsffile{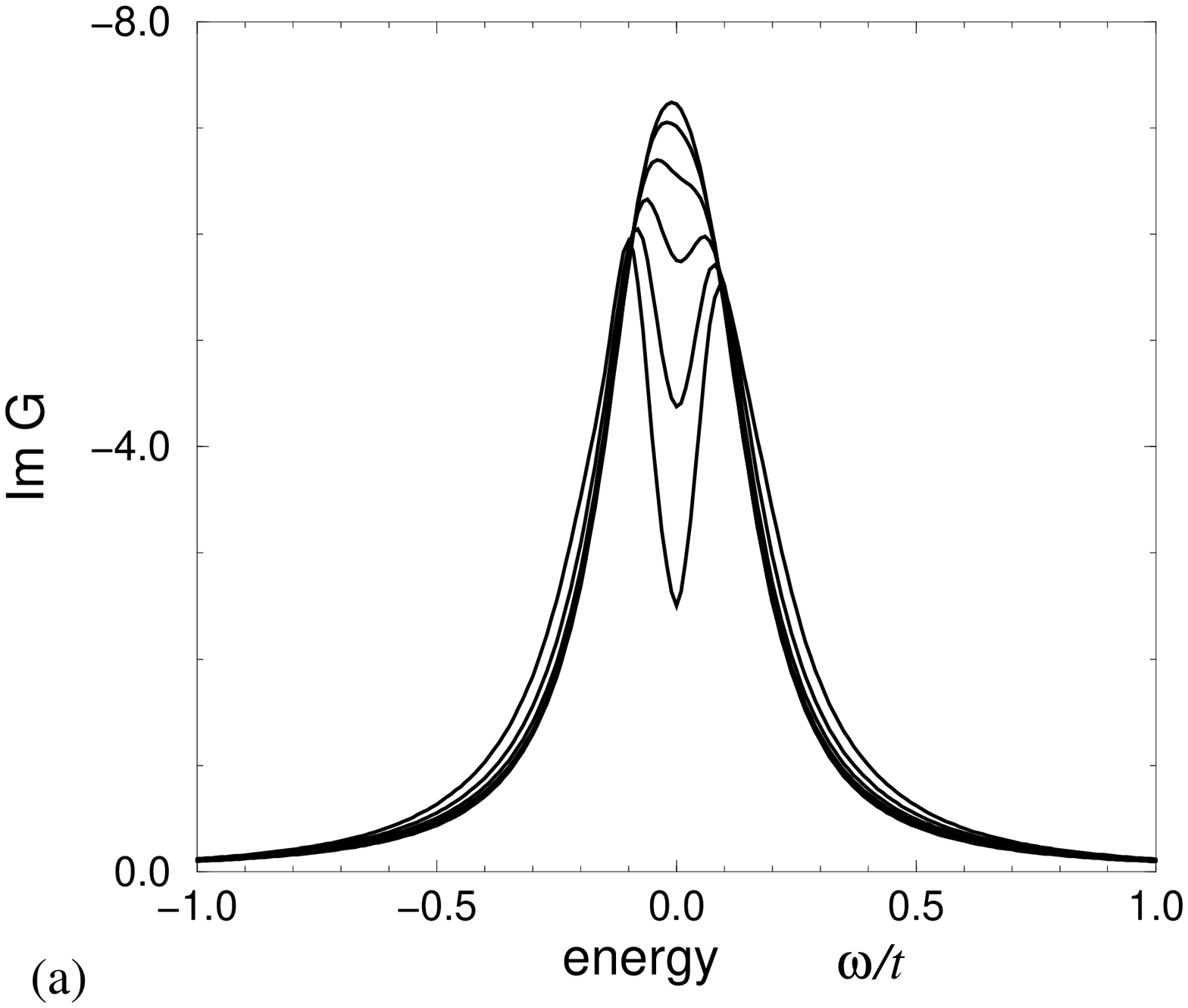}$$
\epsfxsize=7.8cm
$$\epsffile{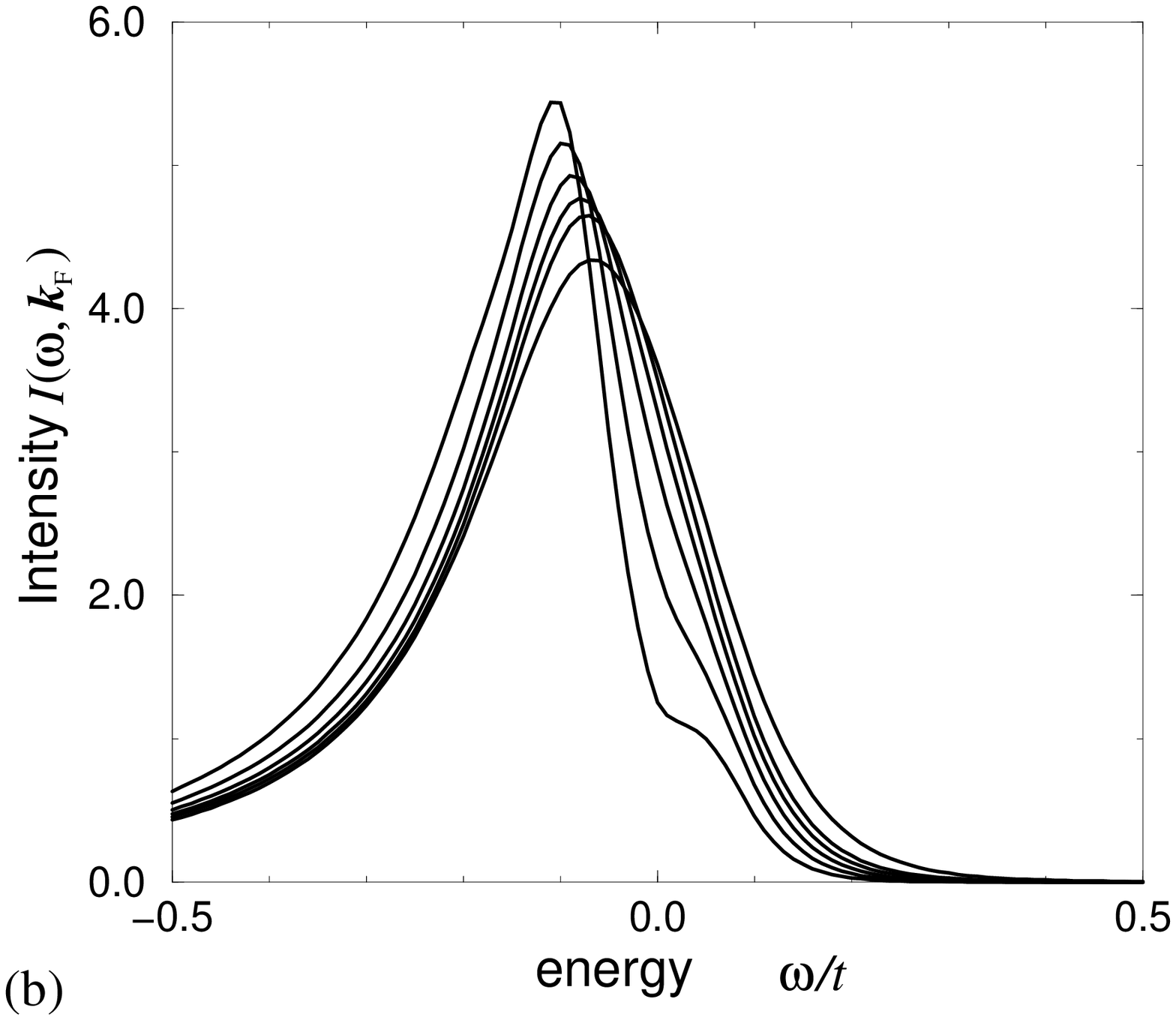}$$
\end{center}
\vspace*{-10mm}
\caption{Temperature variations of (a) the calculated imaginary part of
  the Green's function and (b) the calculated ARPES intensity for
  ${\mib k}_{\rm F}=(\pi,3\pi/64)$ for $\varphi=1$. This momentum
  point is on the Fermi surface and the closest to $(\pi,0)$ in our
  calculation.
  Temperatures in the plotted data are $0.102$, $0.078$, $0.069$,
  $0.06$, $0.051$, and $0.042$ in the energy unit of $t$
  from the data with larger intensity at $\omega=0$ both for (a) and 
  (b).}
\label{FIG_p16403}
\end{figure}
%
Figure~\ref{FIG_p16403}(a) shows $\Ima G(\omega,{\mib k}_{\rm F})$
with ${\mib k}_{\rm F}$ near the flat spot at various temperatures
for $\varphi=1$.
For $T\ge 0.102t$, we have a peak at $\omega=0$,
though it is damped by thermal fluctuations.
At lower temperatures still above $\TPG$, only the low-energy spectral
weights gradually start decreasing.
We note that low-energy fluctuations of $d$SC-SRO grow more
rapidly than those of AFM-SRO below $\TPG$ ($\sim 0.06t$) but above
$T_*$ ($\sim 0.02t$)~\cite{AFM-dSC}. They suppress only
the low-energy part of the peak in the spectral weights.
Well below $\TPG$, $\Ima G$ shows further loss of weights
around $\omega=0$ and the shift to higher energies.
For the same momentum ${\mib k}_{\rm F}$,
we plot the intensities
$I(\omega,{\mib k}_{\rm F})=\Ima G(\omega,{\mib k}_{\rm F}) f(\omega)$
to be observed in ARPES in Fig.~\ref{FIG_p16403}(b), where $f$ is
the Fermi function.
The energy of the midpoint is nearly zero at $T=0.06t(\sim\TPG)$.
For $T<\TPG$, the midpoint shifts to higher binding energies.
This shift amounts to $0.045t\sim 11$meV at $T=0.042t(\sim122{\rm K})$,
in agreement with the experimental situations~\cite{photoemission}

\begin{figure}[tb]
\begin{center}
\epsfxsize=7.8cm
$$\epsffile{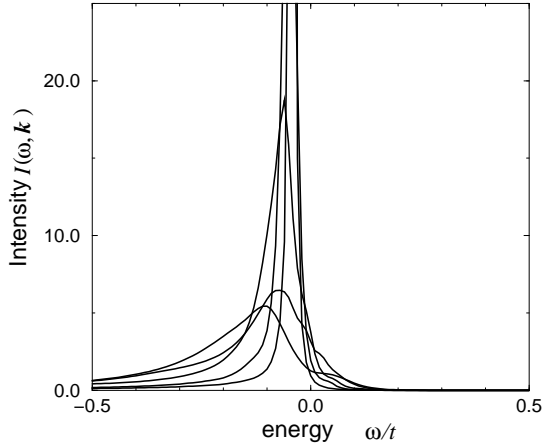}$$
\end{center}
\vspace*{-10mm}
\caption{Calculated ARPES intensity for various momenta
  ${\mib k}\sim{\mib k}_{\rm F}$ at $T=0.42t$ in the $\varphi=1$ case.
  From the sharper side, ${\mib k}=(\frac{\pi}{2},\frac{31}{64}\pi)$,
  $(\frac{40}{64}\pi,\frac{23}{64}\pi)$, $(\frac{48}{64}\pi,\frac{15}{64}\pi)$, $(\frac{56}{64}\pi,\frac{8}{64}\pi)$ and $(\pi,\frac{3}{64}\pi)$.}
\label{FIG_p1arpes07}
\end{figure}
%
Next we discuss the momenum dependence. We plot
$I(\omega,{\mib k})$
for ${\mib k}$'s nearest to the Fermi surface at $T=0.042t$,
in Fig.~\ref{FIG_p1arpes07}.
It shows that the low-energy part of the single-particle excitations
closer to the flat spot is under a stronger suppression,
while those closer to the nodes are better understood as quasiparticles. 
From Fig.~\ref{FIG_p1arpes07}, we can see that the midpoint shift
nearly vanishes at ${\mib k}=(7\pi/8,\pi/8)$.
For momenta closer to the gap nodes, one-peak features are recovered.
These results are consistent with the ARPES data
which suggest the pseudogap developing in the flat shoal region
and the gradual evolution into the $d$SC gap with decrease in $T$.

\begin{figure}[tb]
\begin{center}
\epsfxsize=8cm
$$\epsffile{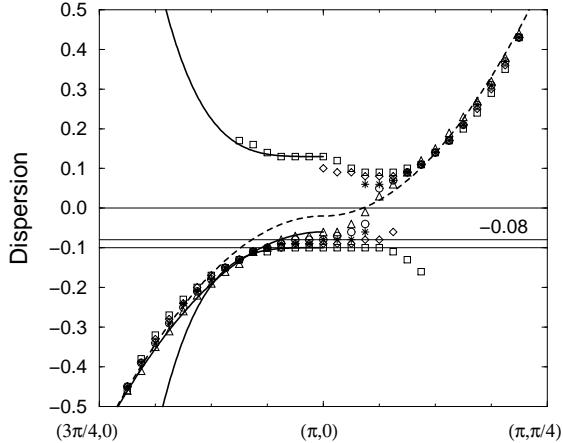}$$
\end{center}
\vspace{-10mm}
\caption{Electronic dispersions scaled by $t$ at $T/t=0.102$ (triangles),
  $0.069$ (circles), $0.06$ (stars), $0.051$ (diamonds) and $0.042$ (squares).
  Solid lines fitted with
  triangles and squares are of the form $k_x^2$ and $k_x^4$,
  respectively. The dashed line represents the bare dispersion.}
\label{FIG_p1dis}
\end{figure}
%
Figure~\ref{FIG_p1dis} shows the ``dispersions'' around the flat spot
compared with the bare one.
We have plotted the energy values at which $\Ima G(\omega,k)$ has
clear peaks for $T/t=0.102$, $0.069$, $0.06$, $0.051$ and $0.042$. 
At $T=0.069t$ slightly higher than $\TPG$,
there exists only one type of ``dispersion'' (circles),
though it jumps from one curve below the chemical potential
to another above that at the remnant Fermi surface.
On the other hand, for $T=0.042t(<\TPG)$,
two types of ``dispersions'' appear due to the precursor effects for
the $d$SC state.
Besides, the dispersion in the direction from $(\pi,0)$ to $(0,0)$
becomes flat and $(k_x-\pi)^4$-like for $|k_x-\pi|\le \pi/8$,
which reminds us of numerical results in a different situation,
doped antiferromagnetic insulator~\cite{Imada98,Assaad98}.
We note that even fermions distant from the remnant Fermi surface
over the flat shoal region have finite spectral weights
at $\omega=0$~\cite{fullpaper}.
It brings about some ambiguity in determining the Fermi surface
around the flat spots.
It also means the breakdown of the Fermi-liquid description
in terms of the quasiparticles,
reflecting the incoherent nature of the flat-spot fermions
due to the quantum fluctuations.

We briefly mention the single-particle properties obtained for $\varphi=0$,
corresponding to the cuprates in the region with no pseudogap
in AFM excitations~\cite{AFM-dSC}. 
In this case, our calculation on $I(\omega,k)$ also shows
no clear pseudogap with a shift of 
midpoint energy on the Fermi surface even around the flap spots.
Instead, at slightly higher temperatures than $T_*$, we have obtained
the precursor to the $d$SC state developing in the spectral weights
at finite positive energies.
For underdoped La${}_{2-x}$Sr${}_x$CuO${}_4$,
where clear spin pseudogap is not observed~\cite{LSCONMR},
the ARPES data~\cite{LSCOARPES}
suggest that the electrons in the diagonal direction
as well as those around the flat spots have rather small weights
at $\omega=0$, in contrast with other cuprates.
They indicate that we have to seriously take into account
the $(\pi/2,\pi/2)$ contributions in considering the susceptibilities
for this compound.

Our results do not reproduce the high-energy backgrounds
observed in ARPES, which extend to even hundreds meV.
Chubukov {\it et al.} have argued that the incoherent backgrounds and
that the peak-dip-hump feature can be obtained from the coupling of
fermions to overdamped spin fluctuations~\cite{Chubukov_PG}.
However, spin fluctuations are underdamped in underdoped cuprates
particularly below $\TPG$.
To obtain the high-energy feature of the single-particle excitations,
it may be necessary to improve the formalism
for calculation of the self-energy in the strong-coupling region.

\section{Summary}

By using the previous solutions of the SCR
which have reproduced the magnetic properties observed in high-$\Tc$
cuprates in the both regions, with and without the pseudogap~\cite{AFM-dSC},
we have calculated the electronic self-energy and the Green's function.

When the mode dampings are suppressed near the transition
($\varphi=1$), $1/T_1T$ shows a decrease below $\TPG(>T_*)$
while the spin correlation length $\xis$ continues to increase until $T_*$.
In this case, our present calculations reproduce
the pseudogap behavior in single-particle excitations,
and the flat and damped dispersion.
Our results show that near $\TPG$, strong $d$SC fluctuations develops
an incoherent feature of single-particle excitations
in the flat shoal region at lower energies than the pseudogap amplitude,
from which the gap-like feature develops below $\TPG$.
The success in reproducing the pseudogap behavior in both of AFM and
single-particle excitations is based on the feedback effects on the
dampings of the collective modes and the competing features 
of low-energy AFM and $d$SC fluctuations.
As concerns the former,
this pseudogap formation in single-particle excitations
around the flat spots qualitatively supports the selfconsistency of
the phenomenological relation (\ref{damp_z=1whole}) used
to obtain the susceptibilities,
though more quantitative analyses are neededx.
The latter leads to the important constraint for the mechanism of the $d$-wave
superconductivity in high-$\Tc$ cuprates:
The repulsion $\usd>0$ is definitely required for reproducing the
pseudogap.
This means that the $d$-wave attraction is not mediated by low-energy
spin fluctuations.
Although it does not necessarily exclude that the attraction might
come from the high-energy part of the spin fluctuations,
it requires a formalism for such incoherent contributions beyond the
conventional weak-coupling approach.

On the other hand, if the dampings are constant ($\varphi=0$),
$\xis$ and $1/T_1T$ reach the maximum values at $T=T_*$.
For this category, we speculate that quasiparticles
around the $(\pi/2,\pi/2)$ point mainly contribute
to the finite damping of AFM fluctuations.
In this case, our present calculations show that the pseudogap region
shrinks in single-particle excitations as well as in AFM excitations,
and that quasiparticles around the flat spots are retained until the
temperature close to $\Tc$, as in the overdoped region.

Assuming the importance of the flat-spot fermions,
we have neglected fermions far from the flat spots
in obtaining the susceptibilities
and discussing the pseudogap formation.
For more detailed discussions of the magnetic and the single-particle
properties in underdoped La${}_{2-x}$Sr${}_x$CuO${}_4$,
it is necessary to consider fermions around $(\pi/2,\pi/2)$ seriously.

To go beyond the present weak-coupling approach,
the direct inclusion of the self-consistent self-energy
corrections in calculating the irreducible susceptibilities is required.
Then the incoherence of the single-particle excitations
near the MIT may seriously modify the AFM and $d$SC susceptibilities,
especially the Curie-Weiss type form
for the dynamic spin susceptibility~\cite{RMP2}.

\acknowledgement
S. O. would like to thank S. Fujiyama for discussions.
The work was supported by "Research for the Future" Program from
the Japan Society for the Promotion of Science under the grant number
JSPS-RFTF97P01103.

\end{document}